\documentclass{article}

\usepackage{graphicx}
\usepackage{amsmath,amssymb}
\usepackage{algorithm}
\usepackage{algpseudocode}
\usepackage{booktabs}
\usepackage{hyperref}

\usepackage{tikz}
\usetikzlibrary{arrows.meta}
\usetikzlibrary{positioning}
\usetikzlibrary{decorations.pathreplacing}
\usetikzlibrary{calc}

\title{Bayesian calibration of adaptive-behavior SIR models for multi-wave COVID-19 incidence in New York City}
\author{Luis A. Barboza \and Carlos Pasquier \and Baltazar Espinoza \and Fabio Sanchez}
\date{May 2025}

\begin{document}

\maketitle

\begin{abstract}
Epidemic incidence reflects both transmission dynamics and adaptive human behavior, yet these mechanisms may be difficult to distinguish from aggregate case data alone. We calibrated four susceptible--infected--recovered (SIR) specifications to weekly confirmed COVID-19 incidence in New York City from June to December 2020, comparing a single continuous SIR trajectory, a wave-initialized SIR model, and two adaptive-behavior models with either shared or wave-specific transmission. Inference was performed using rejection Approximate Bayesian Computation (ABC), and in-sample reconstruction was assessed using root mean squared error (RMSE) and the weighted interval score (WIS). Reinitializing the epidemic state by wave produced the largest structural improvement over the continuous SIR trajectory, reducing mean-based RMSE by 48.6\% and WIS by 17.6\%. Adding delayed prevalence-dependent behavioral adaptation with shared transmission further reduced mean-based RMSE by 23.4\%, but yielded essentially unchanged WIS relative to the wave-initialized SIR model. Allowing transmission to vary by wave did not provide a consistent additional advantage and produced strongly asymmetric posterior-simulation trajectories. Behavioral sensitivity, response midpoint, and delay remained only weakly to partially identified. The clearest posterior structure was a negative association between transmission intensity and the behavioral midpoint, indicating that higher transmission could be compensated by behavioral responses activated at lower prevalence. Sensitivity to ordered behavioral priors further showed that reconstruction and behavioral inference depend materially on structural prior assumptions. These results suggest that adaptive mechanisms can improve multi-wave incidence reconstruction, while aggregate incidence alone is insufficient to sharply separate transmission from behavioral adaptation.
\end{abstract}


\section{Introduction}

Epidemic dynamics reflect more than pathogen transmission alone. They arise
from a coupled system in which infection risk, public information,
institutional interventions, and individual behavior can continually modify one
another. Classical susceptible--infected--recovered (SIR) models, by contrast,
represent transmission through a fixed contact process governed by infection
and recovery \cite{kermack1927,hethcote2000}. Although this simplification is
analytically useful, the COVID-19 pandemic made its limitations particularly
visible: mobility, masking, social contact, and adherence to
non-pharmaceutical interventions changed substantially in response to perceived
risk and epidemic conditions, thereby altering effective transmission over time
\cite{perra2021}.

A substantial literature has therefore incorporated adaptive human behavior
into epidemic models. Behavioral responses have been represented as functions
of current or perceived prevalence, awareness transmitted through social
contact, media and public-health information, or strategic responses to
infection risk \cite{funk2010,funk2009,fenichel2011,reluga2010,verelst2016}.
When such responses feed back into contact rates, they can alter epidemic
timing and generate plateaus, delayed peaks, oscillations, or successive waves
that cannot be attributed to pathogen dynamics alone
\cite{eksin2019,weitz2020}. Delays between changes in epidemiological
conditions and behavioral responses provide an additional mechanism through
which this feedback can shape epidemic trajectories. Mahmud et al.\
\cite{mahmud2025}, for example, introduced a delayed prevalence-dependent
contact-reduction function capable of generating recurrent epidemic waves
through endogenous behavioral adaptation.

The same coupling that makes adaptive models epidemiologically attractive
also creates an inferential problem. Aggregate incidence does not directly
reveal whether a change in epidemic growth was produced by transmission,
behavioral adaptation, or some combination of the two. A model may compensate
for higher transmission with stronger or earlier contact reduction, while a
model that omits behavioral change may absorb behavioral effects into its
transmission coefficient \cite{eksin2019,weitz2020}. Consequently, fitting an
adaptive model successfully does not imply that its behavioral parameters are
well identified. This distinction between trajectory reconstruction and
mechanistic identifiability is particularly important when the fitted outcome
is reported case incidence, because changes in ascertainment may introduce an
additional source of variation that is not explicitly represented in the
epidemic model.

Multi-wave epidemics provide a useful setting in which to examine this
problem because they require a model to accommodate substantial changes in
incidence over time. New York City offers a well-documented example. Following
the severe initial COVID-19 outbreak in spring 2020, the city experienced a
period of phased reopening followed by renewed epidemic growth later in the
year, while public-health policies and population behavior continued to evolve
\cite{nycdata2025a,nycdata2025b,perra2021}. We focus on weekly confirmed cases
from June 7 through December 12, 2020. This interval captures several distinct
changes in the observed incidence trajectory while ending immediately before
COVID-19 vaccination began in the city, allowing the analysis to concentrate
on a pre-vaccination period without introducing an explicit vaccination
compartment.

We use this setting to examine how progressively more flexible SIR
specifications reconstruct multi-wave incidence and what their fitted
parameters can legitimately reveal about behavioral adaptation. Building on
the delayed logistic contact-reduction mechanism of Mahmud et al.\
\cite{mahmud2025}, we compare four models. The first is a classical SIR model
initialized once and propagated continuously through the full study period.
The second retains the non-adaptive SIR structure but independently initializes
three fixed epidemic-wave windows. The remaining two models add delayed
prevalence-dependent contact adaptation to these wave-specific windows, with
transmission either shared across waves or allowed to vary by wave. This
sequence of specifications allows us to distinguish the reconstruction gained
from wave-specific initialization from that associated with an explicit
behavioral mechanism, and to examine whether additional flexibility in
transmission changes inference on the behavioral parameters.

Because calibration of these nonlinear dynamical models is performed without
an explicit observation likelihood, we use rejection Approximate Bayesian
Computation (ABC), a simulation-based approach for inference when direct
likelihood evaluation is unavailable or deliberately avoided
\cite{beaumont2002,csillery2010,toni2009}. Model reconstruction is compared on
a common set of weekly observations using root mean squared error (RMSE) for
central trajectories and the weighted interval score (WIS) to summarize the
accuracy and sharpness of intervals induced by uncertainty in the
ABC-accepted parameter distribution \cite{gneiting2007,bracher2021}. These
comparisons are explicitly in sample and are used descriptively rather than as
out-of-sample forecasting or complexity-adjusted model-selection criteria.

Our analysis addresses three related questions. First, how much of the
multi-wave incidence pattern can be reconstructed by relaxing the requirement
of a single uninterrupted SIR trajectory? Second, after accounting for
wave-specific initialization, does a delayed adaptive-behavior mechanism
provide additional improvement in reconstruction? Third, and most importantly,
to what extent can aggregate incidence distinguish the parameters governing
transmission from those governing prevalence-dependent behavioral adaptation?
We examine the last question through prior-to-posterior updating, posterior
dependence among transmission and behavioral parameters, and sensitivity to
both the ABC acceptance threshold and monotonic structural assumptions on the
behavioral response. In this way, the objective is not only to determine
whether adaptive behavior can reproduce multi-wave epidemic trajectories, but
also to establish what the available incidence data can and cannot identify
about the behavioral mechanisms used to explain them.

\section{Data}

The primary data source was the \textit{COVID-19: Data Trends and Totals}
portal maintained by the New York City Department of Health and Mental
Hygiene, together with its underlying public repository
\cite{nycdata2025a,nycdata2025b}. The dataset reports daily counts of
confirmed and probable COVID-19 cases, hospitalizations, and deaths for New
York City, with additional geographic disaggregation by borough. Events are
indexed by \verb|date_of_interest| rather than by the date on which they were
reported, and the complete daily series is publicly available in
machine-readable form.

We used \verb|CASE_COUNT| as the observed incidence series. This variable
records persons confirmed as COVID-19 cases according to their diagnosis date.
Only confirmed cases were included in the calibration. Over the study period,
the corresponding probable-case series contained 31,324 cases, approximately
23.8\% of the confirmed-case total. We excluded probable cases because they
were defined through a separate ascertainment category and combining the two
series would introduce an additional observation process into a model that does
not explicitly represent case ascertainment. Hospitalizations and deaths were
also not used in the present incidence-only calibration.

The analysis covered June 7 through December 12, 2020. This interval begins
after the intense spring epidemic and around the beginning of New York City's
phased reopening, while ending immediately before COVID-19 vaccination began
in the city on December 14, 2020 \cite{reuters2020,lambert2020}. Restricting
the analysis to this pre-vaccination interval avoids introducing an additional
immunity mechanism that is absent from the SIR models considered here.

Daily confirmed cases were aggregated into 27 consecutive Sunday--Saturday
weekly totals. Weekly aggregation reduces the pronounced day-of-week variation
present in the diagnosis-date series and matches the one-week temporal
resolution of the discrete epidemic models. As a descriptive diagnostic, the
ratio of daily confirmed cases to their centered seven-day mean averaged
approximately 0.63 on Saturdays and Sundays and 1.14 from Monday through
Friday over the study window. Aggregating over complete calendar weeks
therefore reduces this short-period variation without requiring additional
smoothing of the epidemic trajectory.

The resulting weekly series contained 131,574 confirmed cases. Weekly counts
ranged from 1,628 cases in the week beginning August 16 to 19,554 cases in the
week beginning December 6, a ratio of approximately 12 between the largest
and smallest observations. The substantial variation in incidence over the
study period motivated the use of a logarithmic discrepancy for ABC
calibration, as described in Section~\ref{sec:methods}, so that the
late-autumn high-incidence weeks did not dominate the lower-incidence summer
period. The fixed partition of this series into three epidemic-wave windows is
described separately in Section~\ref{sec:methods}; the wave boundaries are a
modeling choice rather than an intrinsic feature of the source data.

\begin{table}[htp]
	\centering
	\begin{tabular}{lr}
		\toprule
		Quantity & Value \\
		\midrule
		Study period                         & June 7--December 12, 2020 \\
		Weekly observations                  & 27 \\
		Total confirmed cases                & 131,574 \\
		Minimum weekly count                 & 1,628 \\
		Maximum weekly count                 & 19,554 \\
		Maximum-to-minimum ratio             & 12.0 \\
		Population, $N$                      & 8,804,190 \\
		\bottomrule
	\end{tabular}
	\caption{Summary of the confirmed COVID-19 case series used in the
		analysis. Daily counts are indexed by diagnosis date and aggregated into
		consecutive Sunday--Saturday weeks. The population is the 2020 Decennial
		Census count for New York City used in the epidemic models.}
	\label{tab:datasummary}
\end{table}

Confirmed cases represent an ascertained subset of infections rather than the
latent infection process represented by the epidemic model. Testing
availability, test-seeking behavior, and the probability that an infection was
diagnosed may have changed during the study period. Because no reporting
fraction or explicit observation model was estimated, such changes cannot be
separated from changes in the fitted epidemic parameters. The model
coefficients should therefore be interpreted as effective parameters calibrated
to confirmed-case incidence rather than as direct estimates of biological
transmission or true infection prevalence.

The source series is also subject to retrospective revision as records are
updated. To ensure internal reproducibility, a fixed extraction of the daily
dataset and the derived weekly series were used consistently across all model
specifications. All analyses use publicly available aggregate data and no
individual-level records. The processed data and analysis code are available at
\texttt{[REPOSITORY LINK]}.

\section{Methods}
\label{sec:methods}

\subsection{Weekly SIR and adaptive-behavior models}

The classical Susceptible--Infected--Recovered (SIR) model partitions a closed population into susceptible ($S$), infectious ($I$), and recovered ($R$) compartments and describes epidemic dynamics through transitions between these states \cite{kermack1927,hethcote2000}. To represent adaptive changes in contact behavior, we use the framework proposed by Mahmud et al.\ \cite{mahmud2025}, in which effective contacts are reduced according to a delayed population-level response to infection prevalence.

In continuous time, the adaptive model can be written as
\begin{align}
	\frac{dS}{dt}
	&=
	-\left[1-r\left(\frac{I(t-\tau)}{N}\right)\right]
	\beta\frac{S(t)I(t)}{N},
	\notag\\
	\frac{dI}{dt}
	&=
	\left[1-r\left(\frac{I(t-\tau)}{N}\right)\right]
	\beta\frac{S(t)I(t)}{N}
	-\gamma I(t),
	\label{eq:modeladapt}\\
	\frac{dR}{dt}
	&=
	\gamma I(t),
	\notag
\end{align}
where $\beta$ denotes transmission intensity, $\gamma$ the recovery rate, $\tau$ the delay between the epidemiological signal and the behavioral response, and $N=S+I+R$ the constant population size. Setting $r\equiv0$ gives the corresponding classical SIR system.

Following \cite{mahmud2025}, the behavioral response is represented by the logistic function
\begin{align}
	r(p)
	=
	\frac{1}{1+\exp[-k(p-c)]},
	\qquad p=\frac{I}{N}.
	\label{eq:adaptativo}
\end{align}
Here $r(p)\in[0,1]$ is interpreted as the proportional reduction in effective contacts at prevalence $p$. The parameter $c$ is the response midpoint, so that $r(c)=0.5$, while $k$ controls the steepness of the prevalence--response relationship. The slope of the response at its midpoint is $k/4$, and larger values of $k$ therefore generate a sharper, more threshold-like transition around $c$. Smaller $k$ values produce a flatter response with weaker dependence on prevalence. In the limit $k\rightarrow0$, $r(p)\rightarrow0.5$ for all $p$; this removes prevalence dependence but does not recover the non-adaptive SIR model, which is defined separately by $r\equiv0$.

The continuous-time formulation provides the mechanistic basis of the model. Because both the observed incidence series and the behavioral delay are resolved at weekly frequency, however, all calibration and inference were performed using a discrete-time formulation with a one-week time step. Specifically, we used a forward-Euler discretization of the continuous system.

For the models fitted to epidemic-wave windows, let $w\in\{1,2,3\}$ index the epidemic wave and let $t$ index the weekly states within that wave. The number of incident cases generated in the transition from week $t$ to week $t+1$ is
\begin{align}
	C_{w,t+1}
	=
	\left[
	1-r_w\left(\frac{I_{w,t-\tau}}{N}\right)
	\right]
	\beta_w\frac{S_{w,t}I_{w,t}}{N},
	\label{eq:weekly_incidence}
\end{align}
and the compartmental states evolve according to
\begin{align}
	S_{w,t+1}
	&=
	S_{w,t}-C_{w,t+1},
	\notag\\
	I_{w,t+1}
	&=
	I_{w,t}+C_{w,t+1}-\gamma I_{w,t},
	\label{eq:weekly_model}\\
	R_{w,t+1}
	&=
	R_{w,t}+\gamma I_{w,t}.
	\notag
\end{align}

For delayed indices preceding the beginning of wave $w$, the infectious state was held at the wave-specific initialization parameter,
\begin{align}
	I_{w,t}=I_w^{\mathrm{init}},
	\qquad t<0.
	\label{eq:history}
\end{align}
The delay was restricted to integer numbers of weeks, consistent with the temporal resolution of the model and data. In specifications with a global transmission coefficient, $\beta_w=\beta$ for all waves; in the non-adaptive models, $r_w(\cdot)\equiv0$.

The single-trajectory SIR benchmark uses the same weekly SIR update with $r\equiv0$, but differs in that it is initialized only once, at the beginning of the full study period, and then propagated continuously through all 27 weeks without resetting $S$, $I$, or $R$ at the wave boundaries.

The population was fixed at $
N=8{,}804{,}190,
$ the 2020 Decennial Census population of New York City. Initial infectious-state counts were treated as continuous model quantities rather than rounded to integers.

No reporting fraction or explicit observation model was estimated. Simulated weekly incidence was compared directly with weekly confirmed-case counts. The fitted compartmental and transmission parameters should therefore be interpreted as effective parameters of a simplified model calibrated to the observed-case process, rather than as direct estimates of true infection prevalence or purely biological transmission rates. In particular, changes in case ascertainment that are not represented explicitly can be absorbed by the fitted parameters.

To preserve compartmental feasibility, the numerical implementation restricts simulated incident infections in a weekly step to the remaining susceptible population and checks conservation of $S+I+R=N$. Activation of this safeguard was tracked as a numerical diagnostic and did not occur among the draws retained in the analyses reported below.

\subsection{Study period, wave partition, and initialization}

The weekly series comprised 27 observations covering June 7--December 12, 2020. For the three wave-based model specifications, the series was partitioned into three non-overlapping epidemic windows. Wave 1 contained eight weekly observations covering June 7--August 1; wave 2 contained eleven observations covering August 2--October 17; and wave 3 contained eight observations covering October 18--December 12.

The wave boundaries were determined through exploratory inspection of the weekly case trajectory, using approximate changes in the epidemic pattern to delimit successive periods. They were not estimated by a changepoint model or defined from an external policy classification. The partition should therefore be viewed as a pragmatic modeling device rather than as a formal identification of epidemiologically unique wave boundaries, and all wave-specific inference is conditional on this fixed partition.

The single-trajectory SIR benchmark does not use the wave boundaries to reset the epidemic state. It is initialized once at the beginning of the study period according to
\begin{align}
	S_0 &= N-I^{\mathrm{init}}, \notag\\
	I_0 &= I^{\mathrm{init}}, \label{eq:single_initial_conditions}\\
	R_0 &= 0, \notag
\end{align}
and the resulting SIR trajectory is propagated continuously through the full 27-week period.

By contrast, the remaining three primary specifications are independently initialized within each epidemic-wave window. We denote by $I_w^{\mathrm{init}}$ the latent infectious-state count used to initialize wave $w$. The initial conditions are
\begin{align}
	S_{w,0} &= N-I_w^{\mathrm{init}}, \notag\\
	I_{w,0} &= I_w^{\mathrm{init}}, \label{eq:wave_initial_conditions}\\
	R_{w,0} &= 0. \notag
\end{align}
For the adaptive models, the pre-initialization history required by the delayed response was held constant at $I_w^{\mathrm{init}}$, as specified in Eq.~\eqref{eq:history}.

For the wave-based models, the first weekly entry of each wave is associated with the initialized compartmental state and with a zero-incidence placeholder rather than with incidence generated by a weekly transition. Consequently, the first observation of each wave was excluded from both the ABC discrepancy and the model-evaluation metrics, leaving 24 post-initialization observations.

For comparability, the single-trajectory benchmark was evaluated on exactly the same set of 24 calendar weeks. Thus, the observations corresponding to the beginnings of waves 2 and 3 were excluded from its discrepancy and evaluation metrics even though they occur internally along the continuous SIR trajectory. These weeks were not removed from the state dynamics: their simulated transitions contribute to the subsequent values of $S$, $I$, and $R$. All four primary specifications were therefore calibrated and evaluated against an identical set of observed weekly counts.

\subsection{Model specifications}

We compared four primary structural specifications that differ in whether the epidemic state is propagated continuously across the full study period or reinitialized by wave, whether behavioral adaptation is included, and whether transmission is shared across waves. The two adaptive models use independent, identically distributed base priors for the wave-specific behavioral parameters in the primary analysis; no monotonicity relationship among waves is imposed. Table~\ref{tab:specs} summarizes the four specifications.

\begin{enumerate}
	\item \textbf{SIR (single trajectory).}
	The non-adaptive benchmark with a single transmission coefficient $\beta$, recovery coefficient $\gamma$, and initialization parameter $I^{\mathrm{init}}$. The model is initialized once at the beginning of the study period and propagated continuously through all 27 weeks. This specification assesses how much of the observed multi-wave incidence pattern can be reconstructed by a single static SIR trajectory without wave-specific resets or behavioral adaptation.
	
	\item \textbf{SIR (wave-specific initialization).}
	The non-adaptive model with global $\beta$ and $\gamma$ but three independently initialized wave windows with separate initialization parameters
	$$
	I_1^{\mathrm{init}},\quad
	I_2^{\mathrm{init}},\quad
	I_3^{\mathrm{init}}.
	$$
	Relative to the single-trajectory benchmark, this specification introduces both wave segmentation and wave-specific reinitialization while retaining a common transmission and recovery structure. The comparison therefore measures the gain obtained by allowing each epidemic window to begin from its own latent infectious state, rather than isolating the effect of a single parameter change.
	
	\item \textbf{Adaptive ($\beta$ global).}
	The adaptive model with global $\beta$, $\gamma$, and $\tau$, together with wave-specific behavioral sensitivities $k_w$, behavioral midpoints $c_w$, and initialization parameters $I_w^{\mathrm{init}}$. In the primary specification, the three $k_w$ values and the three $c_w$ values are sampled independently from common base priors, with no ordering imposed across waves.
	
	\item \textbf{Adaptive ($\beta$ variable).}
	The adaptive model with wave-specific transmission coefficients $\beta_w$, behavioral parameters $k_w$ and $c_w$, and initialization parameters $I_w^{\mathrm{init}}$, while $\gamma$ and $\tau$ remain global. This specification tests whether additional wave-specific flexibility in transmission materially changes reconstruction or inference on the behavioral component.
\end{enumerate}

\begin{table}[htp]
	\centering
	\begin{tabular}{lccc}
		\toprule
		Model & Global parameters & Wave-specific parameters & Free parameters \\
		\midrule
		SIR (single trajectory)
		& $\beta,\gamma,I^{\mathrm{init}}$
		& none
		& 3 \\
		SIR (wave-specific initialization)
		& $\beta,\gamma$
		& $I_w^{\mathrm{init}}$
		& 5 \\
		Adaptive ($\beta$ global)
		& $\beta,\gamma,\tau$
		& $k_w,c_w,I_w^{\mathrm{init}}$
		& 12 \\
		Adaptive ($\beta$ variable)
		& $\gamma,\tau$
		& $\beta_w,k_w,c_w,I_w^{\mathrm{init}}$
		& 14 \\
		\bottomrule
	\end{tabular}
	\caption{Structure of the four primary model specifications. The single-trajectory SIR is propagated continuously over the full study period, whereas the remaining three specifications use independently initialized epidemic-wave windows. Free-parameter counts include all global and wave-specific quantities sampled in the ABC calibration. The adaptive models are unrestricted with respect to ordering of $k_w$ and $c_w$ in the primary analysis.}
	\label{tab:specs}
\end{table}

\subsection{Priors and ABC calibration}

We assigned broad priors designed to span a wide range of weekly epidemic and behavioral dynamics rather than to encode precise external estimates. Parameters whose supports span multiple orders of magnitude were sampled uniformly on the $\log_{10}$ scale, whereas $\beta$ and $\gamma$ were sampled uniformly on their natural weekly scale. Table~\ref{tab:priors} gives the complete specification.

The support $\beta\in[0.4,3.0]$ allows substantial variation in weekly transmission intensity, while $\gamma\in[0.05,0.6]$ allows between $5\%$ and $60\%$ of the current infectious compartment to leave that compartment during a weekly update. These ranges were intentionally broad because $\beta$ and $\gamma$ are effective coefficients of a simplified weekly model fitted to reported cases rather than externally fixed biological quantities.

For the single-trajectory benchmark, the initialization parameter was assigned
$$
\log_{10}I^{\mathrm{init}}\sim U(2.8,4.8).
$$
For the wave-based models, each independently estimated initialization parameter uses the corresponding base prior
$$
\log_{10}I_w^{\mathrm{init}}\sim U(2.8,4.8).
$$
This support corresponds to approximately $6.3\times10^2$ to $6.3\times10^4$ individuals, or approximately $0.007\%$--$0.72\%$ of the fixed New York City population, and permits substantial variation in the latent infectious state used to initialize a trajectory.

For the behavioral parameters,
$$
\log_{10}k\sim U(2,5)
$$
spans responses ranging from relatively gradual to sharply threshold-like, while
$$
\log_{10}c
\sim
U\!\left(-4,\log_{10}(5\times10^{-2})\right)
$$
allows the $50\%$ response midpoint to range from $0.01\%$ to $5\%$ prevalence. The delay was assigned a discrete uniform prior on
$$
\tau\in\{0,1,2,3,4\}\ \text{weeks},
$$
covering responses from no additional behavioral delay to approximately one month. Wave-specific copies of a parameter use the same base prior.

\begin{table}[htp]
	\centering
	\begin{tabular}{llll}
		\toprule
		Parameter & Interpretation & Support & Sampling scale \\
		\midrule
		$\beta$
		& weekly transmission coefficient
		& $[0.4,3.0]$
		& natural \\
		$\gamma$
		& weekly recovery coefficient
		& $[0.05,0.6]$
		& natural \\
		$\tau$
		& behavioral delay (weeks)
		& $\{0,1,2,3,4\}$
		& discrete \\
		$k$
		& behavioral sensitivity
		& $[10^2,10^5]$
		& $\log_{10}$ \\
		$c$
		& behavioral midpoint (prevalence)
		& $[10^{-4},5\times10^{-2}]$
		& $\log_{10}$ \\
		$I^{\mathrm{init}}$
		& initial infectious-state count
		& $[10^{2.8},10^{4.8}]$
		& $\log_{10}$ \\
		\bottomrule
	\end{tabular}
	\caption{Base prior specifications used for ABC calibration. Priors are uniform over the stated support on the indicated sampling scale. Wave-specific quantities, including $I_w^{\mathrm{init}}$, use independent copies of the corresponding base prior in the primary analysis.}
	\label{tab:priors}
\end{table}

We used rejection Approximate Bayesian Computation (ABC) to calibrate the deterministic weekly simulators without specifying an explicit observation likelihood \cite{beaumont2002,csillery2010,toni2009}. For each model, candidate parameter vectors were drawn from the corresponding prior. The single-trajectory SIR was simulated continuously over the full 27-week period, whereas the remaining three specifications were simulated over their independently initialized epidemic-wave windows.

The resulting weekly incidence was compared with the observed confirmed-case series using the same 24 calendar weeks for every model. The discrepancy was defined as the mean squared error on the $\log(1+x)$ scale,
\begin{align}
	d\bigl(\widehat{C}^{(i)},Y\bigr)
	=
	\frac{1}{|\mathcal{T}|}
	\sum_{t\in\mathcal{T}}
	\left[
	\log\left(1+\widehat{C}^{(i)}_{t}\right)
	-
	\log\left(1+Y_{t}\right)
	\right]^2,
	\label{eq:discrepancy}
\end{align}
where $\mathcal{T}$ contains the 24 retained calendar weeks. Every retained week contributes equally to the discrepancy. The logarithmic transformation reduces the dominance of the largest weekly counts while retaining sensitivity to relative differences during the lower-incidence portions of the study period.

For each primary model specification, we generated $
M=500{,}000
$ candidate parameter vectors. Rather than fixing an absolute ABC tolerance in advance, candidates were ranked by discrepancy. The primary analysis retained the lowest-discrepancy $
q=0.005
$
fraction, corresponding to 2,500 accepted draws. The effective tolerance $\varepsilon$ for each model is therefore the largest discrepancy among its 2,500 retained candidates and is model-specific. The retained draws constitute the empirical ABC approximation used for parameter summaries and posterior simulations.

Algorithm~\ref{alg:abc} summarizes the calibration procedure.

\begin{algorithm}[htp]
	\caption{ABC rejection calibration used for the primary model specifications}
	\label{alg:abc}
	\begin{algorithmic}[1]
		\State \textbf{Input:} observed weekly cases $Y$, model-specific prior $\pi(\theta)$,
		number of proposals $M=500{,}000$, acceptance proportion $q=0.005$
		\For{$i=1,\ldots,M$}
		\State Draw $\theta^{(i)}\sim\pi(\theta)$
		\State Simulate either the full 27-week trajectory or the three wave windows, according to the model specification
		\State Retain simulated incidence for the 24 common calendar weeks
		\State Compute $d^{(i)}=d(\widehat{C}^{(i)},Y)$ using Eq.~\eqref{eq:discrepancy}
		\EndFor
		\State Rank the $M$ proposals by $d^{(i)}$
		\State Retain the $qM=2{,}500$ proposals with the smallest discrepancies
		\State Set $\varepsilon$ to the largest discrepancy in the retained set
		\State \textbf{Return:} the 2,500 retained parameter vectors
	\end{algorithmic}
\end{algorithm}

\subsection{Sensitivity analyses}

We considered two sources of sensitivity: the ABC acceptance threshold and structural assumptions imposed on the behavioral priors.

First, sensitivity to ABC tolerance was assessed by forming nested accepted sets corresponding to
$$
q\in\{0.0025,0.005,0.01\},
$$
that is, the lowest-discrepancy 0.25\%, 0.5\%, and 1\% of each pool of 500,000 proposals. These thresholds yield 1,250, 2,500, and 5,000 accepted draws, respectively. The 0.5\% set is used for the primary results, while the other two sets assess whether substantive conclusions depend on a tighter or looser ABC approximation.

Second, both adaptive models were re-estimated under ordered behavioral priors. The primary models allow $k_w$ and $c_w$ to vary independently across waves because the incidence data do not establish a direction of behavioral change a priori. Nevertheless, a progressively attenuated prevalence-dependent response across successive waves is substantively plausible in this application. The restriction
$$
k_1\ge k_2\ge k_3
$$
requires the maximum slope of the logistic response, $k_w/4$, not to increase across successive waves and therefore represents a progressively less steep response to changes in prevalence. The complementary restriction

$$
c_1\le c_2\le c_3
$$
requires the prevalence associated with a $50\%$ contact reduction not to decrease, representing a progressively higher prevalence threshold for the same midpoint response.

Taken jointly, these restrictions encode a structured hypothesis of progressively attenuated prevalence-responsive behavior over successive epidemic waves. Such a pattern could be compatible with mechanisms such as habituation to epidemic risk, declining salience of case increases, or greater tolerance for maintaining contacts as the pandemic progressed. These mechanisms are not observed directly in the present data, however, and the ordered specification was therefore treated as a sensitivity analysis rather than as the primary behavioral model.

Computationally, the ordered prior was generated by first drawing three independent values from the same base prior for each behavioral parameter and then sorting the $k$ values in decreasing order and the $c$ values in increasing order. The ordering therefore changes the wave-specific marginal priors before the data are observed. Consequently, an ordered posterior trajectory cannot by itself be interpreted as empirical evidence for monotonic behavioral change. We instead use these runs to determine how imposing this substantively motivated structure affects model reconstruction and posterior inference.

\subsection{Posterior simulation and model evaluation}

For each accepted ABC draw, the corresponding deterministic model was re-simulated according to its structural specification. The single-trajectory SIR was propagated continuously over all 27 weeks, whereas the remaining three models were re-simulated over their independently initialized epidemic-wave windows. The distribution of trajectories across accepted draws was summarized by the posterior-simulation mean, median, and central $50\%$, $90\%$, and $95\%$ intervals. These intervals propagate uncertainty in the ABC-accepted parameter distribution only. Because the model contains no stochastic observation component, they do not represent full predictive intervals for variability in reported case counts.

All evaluation metrics were computed on the same 24 calendar weeks used in the ABC discrepancy. For the single-trajectory SIR, the omitted observations at the wave boundaries remain part of the simulated state evolution even though their observed case counts do not contribute directly to the evaluation metrics.

Point reconstruction was assessed using root mean squared error (RMSE) on the natural scale of weekly cases. We computed RMSE separately for the posterior-simulation mean and median,
\begin{align}
	\mathrm{RMSE}_{s}
	=
	\left[
	\frac{1}{|\mathcal{T}|}
	\sum_{t\in\mathcal{T}}
	\left(
	Y_t-\widehat{C}^{(s)}_t
	\right)^2
	\right]^{1/2},
	\qquad
	s\in\{\mathrm{mean},\mathrm{median}\}.
	\label{eq:rmse}
\end{align}
Reporting both summaries reduces dependence of the comparison on a single measure of central tendency when the posterior-simulation distributions are asymmetric.

We also used the weighted interval score (WIS) to summarize interval performance \cite{gneiting2007,bracher2021}. For a central $(1-\alpha)$ interval with lower and upper endpoints $l$ and $u$, the interval score is
\begin{align}
	\mathrm{IS}_{\alpha}(l,u;y)
	=
	(u-l)
	+
	\frac{2}{\alpha}(l-y)\mathbf{1}\{y<l\}
	+
	\frac{2}{\alpha}(y-u)\mathbf{1}\{y>u\},
	\label{eq:is}
\end{align}
which combines interval width with penalties for observations falling below or above the interval. Given $K$ central intervals and median $m$, the WIS is
\begin{align}
	\mathrm{WIS}(F,y)
	=
	\frac{1}{K+1/2}
	\left[
	\frac{1}{2}|y-m|
	+
	\sum_{j=1}^{K}
	w_j
	\mathrm{IS}_{\alpha_j}(l_j,u_j;y)
	\right],
	\qquad
	w_j=\frac{\alpha_j}{2}.
	\label{eq:wis}
\end{align}
We used the $K=3$ central intervals at $50\%$, $90\%$, and $95\%$, corresponding to
$$
\alpha\in\{0.50,0.10,0.05\},
$$
and averaged the weekly WIS over the 24 retained observations. Lower values indicate better combined interval accuracy and sharpness.

Because WIS is formally a proper scoring rule for predictive distributions, its interpretation here requires qualification. In this analysis it is applied to the distribution of trajectories induced by ABC parameter uncertainty alone, without an observation-level stochastic model. We therefore use WIS as a descriptive measure of in-sample interval performance rather than as a formal assessment of out-of-sample probabilistic forecast calibration.

As complementary diagnostics, we computed empirical coverage of the central $50\%$, $90\%$, and $95\%$ posterior-simulation intervals over the same 24 observations. These coverage values were interpreted descriptively and jointly with interval width and WIS.

Finally, all reported comparisons are in sample: the same weekly series is used for ABC calibration and model evaluation. RMSE, WIS, and coverage therefore assess reconstruction of the fitted data rather than out-of-sample forecasting skill. Moreover, the four specifications differ substantially in parameter dimension and structural flexibility, and the reported metrics contain no explicit penalty for model complexity. Differences among models should consequently be interpreted as descriptive comparisons of fitted reconstruction and parameter-induced interval performance, rather than as complexity-adjusted model-selection criteria.

\section{Results}

\subsection{In-sample reconstruction across primary models}

\begin{figure}[htp]
	\begin{center}
		\begin{tabular}{c}
			\includegraphics[scale=0.35]{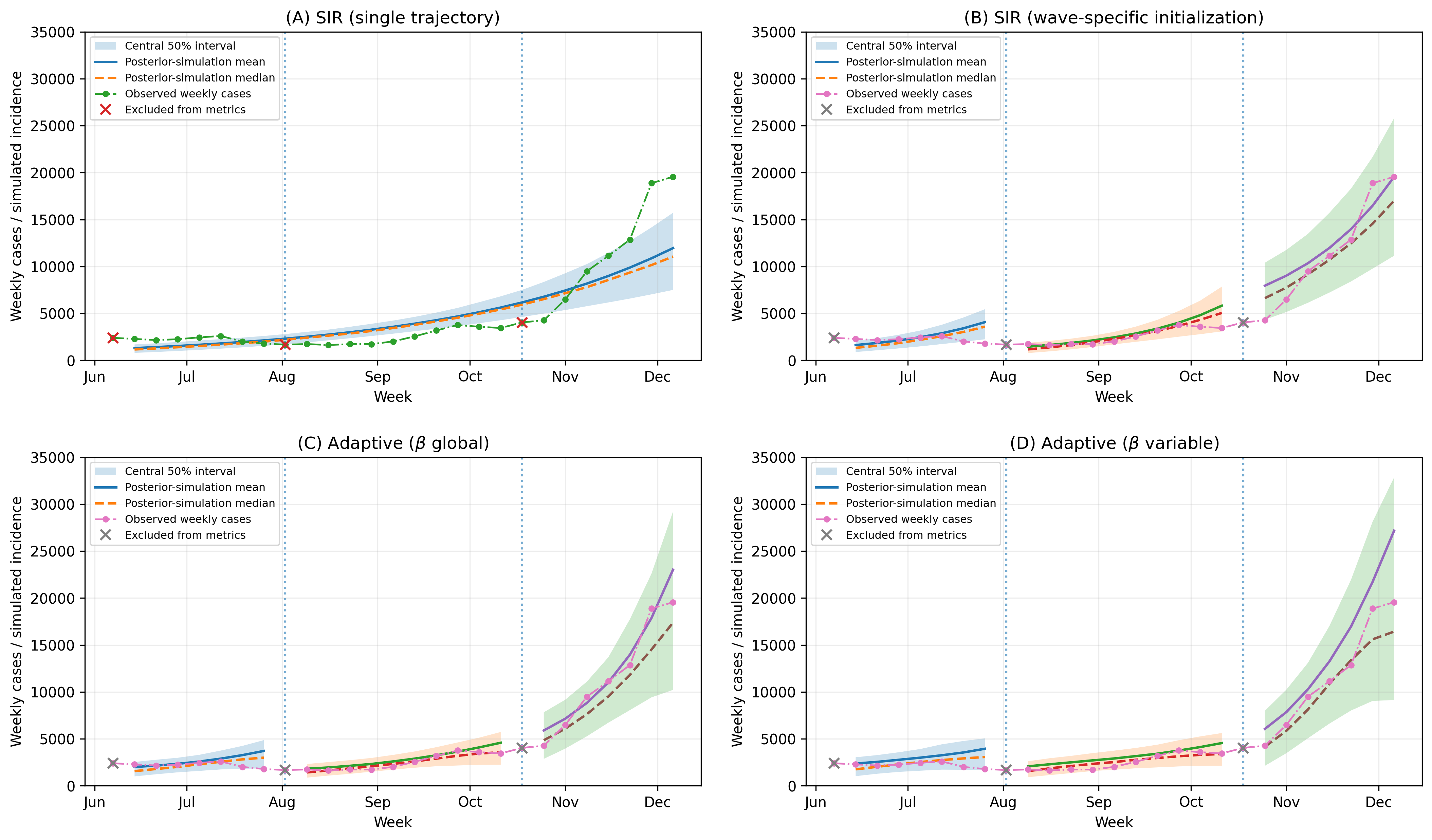}
		\end{tabular}		
	\end{center}
	\caption{Posterior-simulation summaries for the four primary model specifications fitted to weekly COVID-19 case incidence in New York City. Panel (A) shows the SIR single-trajectory benchmark, panel (B) the SIR model with wave-specific initialization, panel (C) the Adaptive ($\beta$ global) model, and panel (D) the Adaptive ($\beta$ variable) model. In each panel, observed weekly incidence is shown together with the posterior-simulation mean, posterior-simulation median, and central 50\% interval. The single-trajectory SIR is propagated continuously across the full 27-week study period, whereas the other three specifications are independently initialized within each epidemic-wave window. Vertical dotted lines indicate the fixed wave boundaries. The three observations excluded from the common set of 24 weeks used for ABC calibration and model-evaluation metrics are marked separately; for the single-trajectory SIR, these observations remain part of the state propagation despite being excluded from scoring.
	}
	\label{fig:ajustes}
\end{figure}

The four primary model specifications differed substantially in their ability to reconstruct the observed weekly case trajectory (Table~\ref{tab:measures}; Figure~\ref{fig:ajustes}). All comparisons reported here use the primary ABC acceptance proportion, $q=0.005$, and the same 24 retained calendar weeks.

\begin{table}[htp]
	\centering
	{\footnotesize
		\begin{tabular}{lrrrrrr}
			\toprule
			Model & RMSE (mean) & RMSE (median) & WIS & Cov. 50\% & Cov. 90\% & Cov. 95\% \\
			\midrule
			SIR (single trajectory)
			& 2638.3
			& 2857.5
			& 781.0
			& 25.0
			& 83.3
			& 83.3 \\
			
			SIR (wave-specific initialization)
			& 1355.5
			& 1326.5
			& \textbf{643.8}
			& 87.5
			& 100.0
			& 100.0 \\
			
			Adaptive ($\beta$ global)
			& \textbf{1038.3}
			& 1209.6
			& 653.1
			& 95.8
			& 100.0
			& 100.0 \\
			
			Adaptive ($\beta$ variable)
			& 2102.6
			& \textbf{1067.7}
			& 831.1
			& 100.0
			& 100.0
			& 100.0 \\
			\bottomrule
		\end{tabular}
	}
	\caption{In-sample reconstruction performance of the four primary model specifications at the main ABC acceptance proportion $q=0.005$. RMSE, WIS, and empirical interval coverage were computed over the same 24 retained calendar weeks. For the single-trajectory SIR, all 27 weeks contributed to state propagation, but the same three observations excluded from scoring in the wave-initialized models were omitted from the evaluation metrics to ensure a common comparison set. The adaptive models use unrestricted wave-specific behavioral priors in the primary analysis. Lower RMSE and WIS indicate better reconstruction. Coverage is reported descriptively and should not be interpreted as out-of-sample predictive calibration.}
	\label{tab:measures}
\end{table}

The SIR single-trajectory benchmark provided the poorest point reconstruction among the simpler SIR specifications, with an RMSE of 2,638.3 cases for the posterior-simulation mean and 2,857.5 for the median. Its WIS was 781.0, while empirical coverage of the central $50\%$, $90\%$, and $95\%$ posterior-simulation intervals was $25.0\%$, $83.3\%$, and $83.3\%$, respectively. As shown in Figure~\ref{fig:ajustes}A, a single SIR trajectory with constant transmission and recovery parameters was unable to reproduce the successive changes in incidence observed across the study period.

Allowing the epidemic state to be reinitialized separately for each wave produced a marked improvement. The SIR model with wave-specific initialization reduced the mean-based RMSE to 1,355.5 and the median-based RMSE to 1,326.5, corresponding to reductions of $48.6\%$ and $53.6\%$, respectively, relative to the single-trajectory benchmark. WIS decreased by $17.6\%$, from 781.0 to 643.8. Coverage also increased to $87.5\%$ for the central $50\%$ interval and $100\%$ for both the $90\%$ and $95\%$ intervals. Thus, the largest structural improvement among the non-adaptive models was associated with replacing a single continuous SIR trajectory by independently initialized epidemic-wave windows.

Adding adaptive behavior while retaining a global transmission coefficient further improved point reconstruction. The Adaptive ($\beta$ global) model achieved the lowest mean-based RMSE among the four primary specifications, 1,038.3 cases, a $23.4\%$ reduction relative to the wave-specific SIR model. Its median-based RMSE was 1,209.6, an $8.8\%$ reduction. In contrast, its WIS was 653.1, only $1.4\%$ higher than the 643.8 obtained by the wave-specific SIR model. The two models therefore differed more clearly in point reconstruction than in interval performance. The adaptive global-$\beta$ specification covered $95.8\%$ of observations with its central $50\%$ interval and all retained observations with its $90\%$ and $95\%$ intervals.

Allowing $\beta$ to vary across waves did not produce a uniform improvement. The Adaptive ($\beta$ variable) model had the lowest median-based RMSE, 1,067.7, but its mean-based RMSE increased to 2,102.6 and its WIS to 831.1. Relative to the Adaptive ($\beta$ global) model, this corresponds to an $11.7\%$ improvement in median-based RMSE but a $102.5\%$ increase in mean-based RMSE and a $27.3\%$ increase in WIS. The large difference between mean- and median-based RMSE indicates substantial asymmetry in the distribution of reconstructed trajectories under this more flexible specification. Its central $50\%$, $90\%$, and $95\%$ intervals covered all 24 retained observations.

Overall, the model comparison identifies two distinct patterns. First, replacing a single continuous SIR trajectory with wave-specific initialization produced a large improvement in reconstruction. Second, adding adaptive behavior with a shared transmission coefficient further improved central trajectory reconstruction, but did not improve WIS relative to the simpler wave-specific SIR model. Additional wave-specific flexibility in $\beta$ did not yield a consistent advantage across the evaluation criteria. Because these metrics are calculated on the same data used for ABC calibration and do not penalize model complexity, these results describe relative in-sample reconstruction rather than out-of-sample forecasting performance or formal model selection.

\subsection{Behavioral-parameter inference and identifiability}

\begin{figure}[htp!]
	\centering
	\includegraphics[scale=0.5]{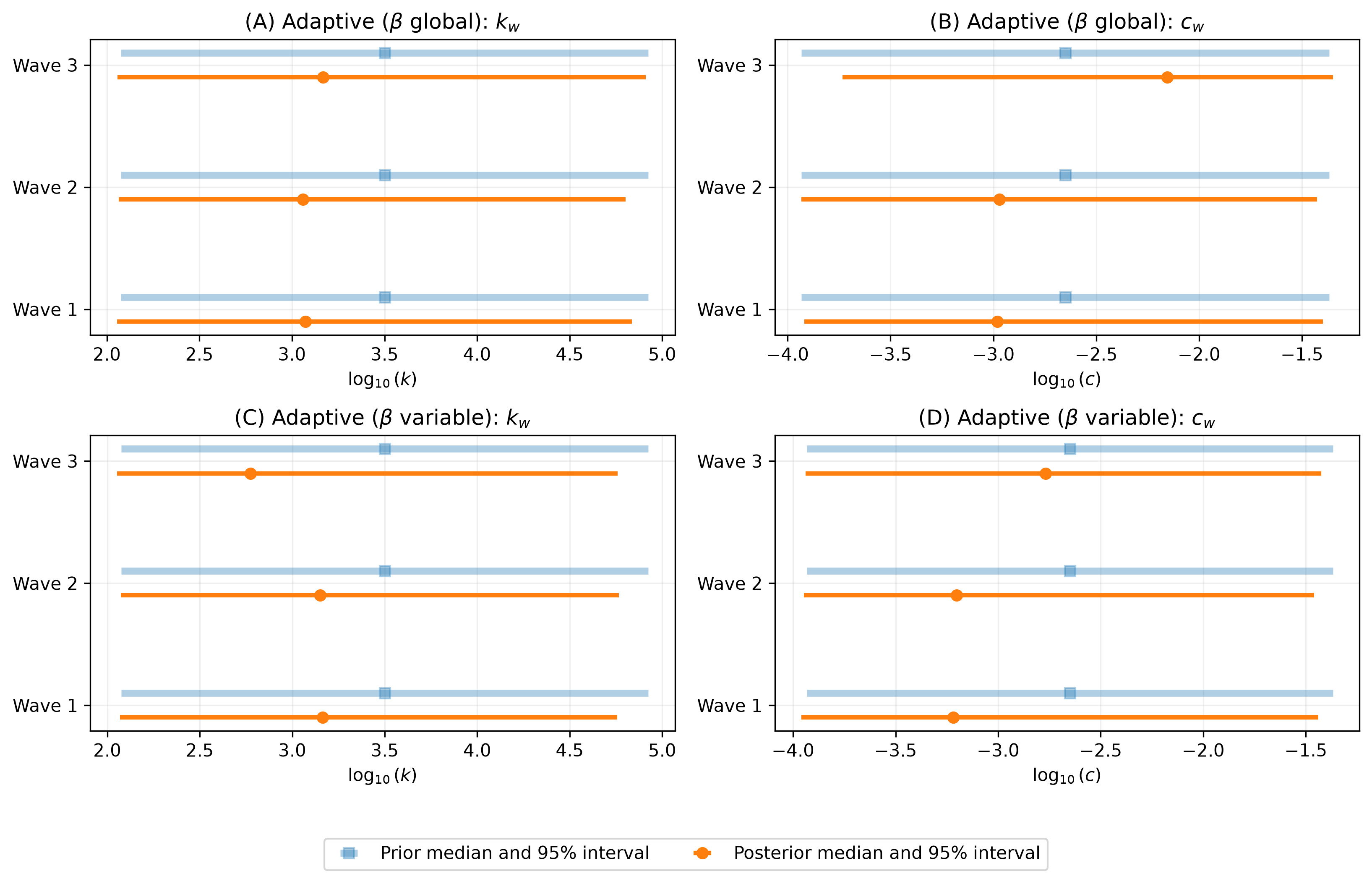}
	\caption{Prior-to-posterior updating of the behavioral parameters in the two primary adaptive model specifications. Panels (A) and (B) show the behavioral sensitivity $k_w$ and response midpoint $c_w$, respectively, for the Adaptive ($\beta$ global) model; panels (C) and (D) show the corresponding quantities for the Adaptive ($\beta$ variable) model. Parameters are displayed on the $\log_{10}$ scale. Points denote prior and posterior medians, and horizontal segments denote the corresponding central 95\% intervals. The limited contraction of most posterior intervals relative to their prior supports indicates weak to partial identification of the behavioral parameters from weekly incidence alone.}
	\label{fig:figure2priorposteriorbehavior}
\end{figure}

Despite the ability of the adaptive specifications to reconstruct the observed incidence trajectories, the behavioral parameters were only weakly to partially identified by the incidence data. Figure~\ref{fig:figure2priorposteriorbehavior} compares the prior and posterior medians and central 95\% intervals of $k_w$ and $c_w$ on the $\log_{10}$ scale. Across both adaptive specifications, the posterior intervals remained close in width to their corresponding prior intervals, with only modest contraction. Thus, good reconstruction of weekly incidence did not translate into precise estimation of the parameters governing the behavioral response.

Under the Adaptive ($\beta$ global) specification, the posterior medians of $k_1$, $k_2$, and $k_3$ were approximately 1,180, 1,147, and 1,476, respectively. There was consequently no descriptive decline in the posterior medians across waves. The posterior probability of the complete ordering
$$
k_1\geq k_2\geq k_3
$$
was 0.183, close to the probability $1/6\approx0.167$ assigned to any specific ordering of three exchangeable continuous quantities under the unrestricted prior. The behavioral midpoints showed somewhat greater differentiation: posterior medians were approximately 0.00104, 0.00107, and 0.00699 for $c_1$, $c_2$, and $c_3$, respectively. Consistent with this shift, the posterior probabilities that $c_3$ exceeded $c_1$ and $c_2$ were approximately 0.77. However, the corresponding posterior intervals remained broad, and $c_1$ and $c_2$ showed essentially no ordering preference. These results are therefore more consistent with a possible upward shift in the third-wave behavioral midpoint than with a well-resolved monotonic progression across all three waves.

Allowing transmission to vary by wave did not eliminate the uncertainty in the behavioral parameters. Under the Adaptive ($\beta$ variable) specification, posterior median values of $k_1$, $k_2$, and $k_3$ were approximately 1,457, 1,413, and 595, respectively. Although this indicates a posterior shift toward lower $k_3$, the central 95\% intervals remained broad and strongly overlapping (Figure~\ref{fig:figure2priorposteriorbehavior}C). The probability of the full ordering $k_1\geq k_2\geq k_3$ was 0.250, while the corresponding probability for $c_1\leq c_2\leq c_3$ was 0.249. Thus, the unrestricted adaptive models provide at most limited support for systematic monotonic changes in behavioral sensitivity or response midpoint across successive waves. Detailed pairwise and joint ordering probabilities are reported in the Supplementary Material.

\begin{figure}
	\centering
	\includegraphics[scale=0.4]{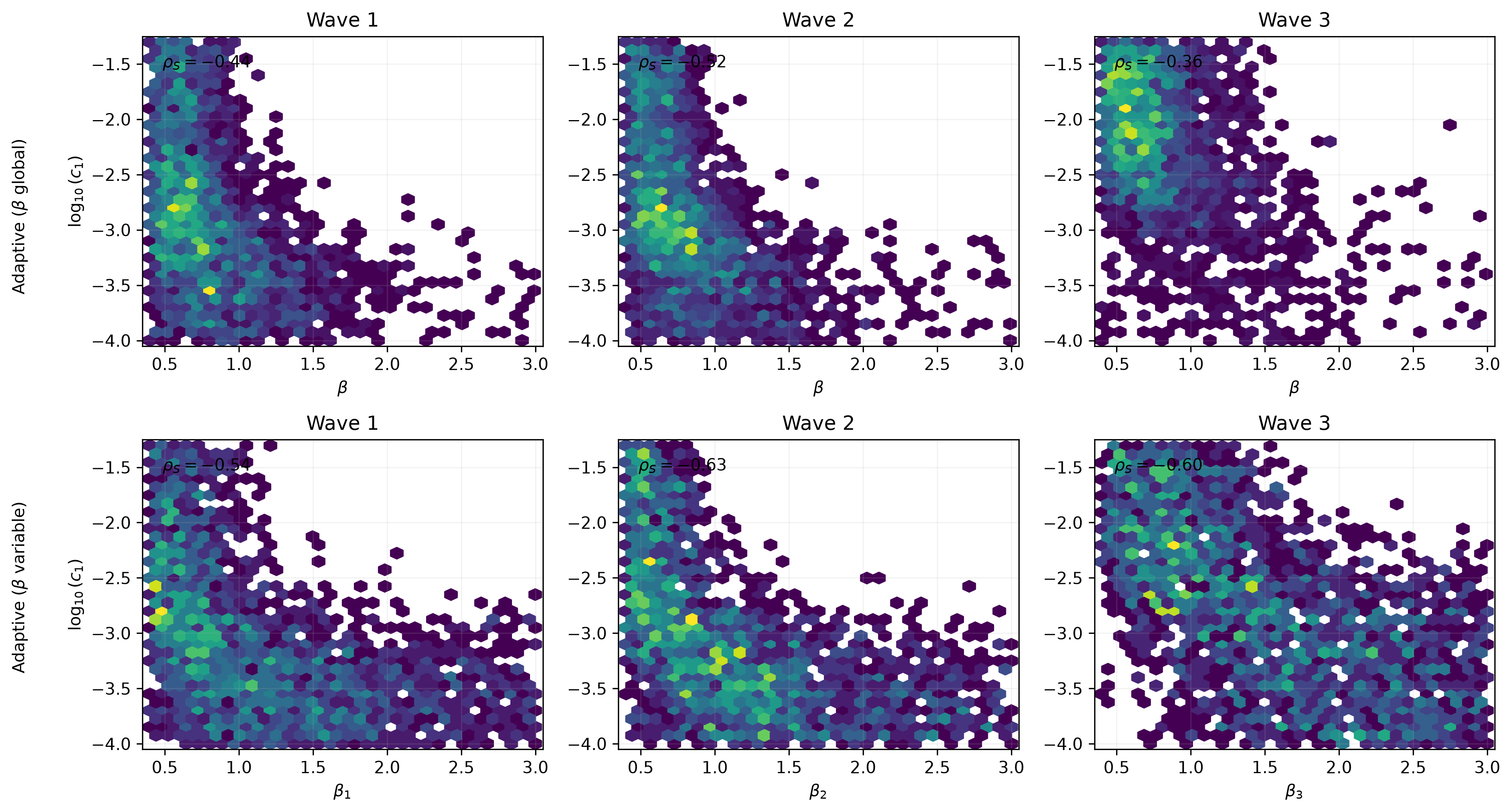}
	\caption{Posterior dependence between transmission intensity and the behavioral response midpoint in the two primary adaptive model specifications. The upper row shows the shared transmission coefficient $\beta$ against $\log_{10}(c_w)$ for waves 1--3 under the Adaptive ($\beta$ global) model. The lower row shows the corresponding wave-specific pairs $\beta_w$ and $\log_{10}(c_w)$ under the Adaptive ($\beta$ variable) model. Hexagonal bins represent the density of the 2,500 ABC-accepted draws at $q=0.005$, and $\rho_S$ denotes the Spearman rank correlation. The consistently negative associations indicate that higher fitted transmission intensity can be offset by a behavioral response reaching its midpoint at lower prevalence, revealing confounding between transmission and behavioral adaptation.}
	\label{fig:figure3betacconfounding}
\end{figure}

A more consistent feature of the posterior distributions was the dependence between transmission intensity and the behavioral midpoint $c_w$ (Figure~\ref{fig:figure3betacconfounding}). In the Adaptive ($\beta$ global) model, the Spearman correlations between the shared transmission coefficient and the wave-specific behavioral midpoint were
$$
\rho_S\!\left(\beta,\log_{10}c_1\right)=-0.44,\qquad
\rho_S\!\left(\beta,\log_{10}c_2\right)=-0.52,\qquad
\rho_S\!\left(\beta,\log_{10}c_3\right)=-0.36.
$$
The negative association was stronger when transmission was allowed to vary by wave:
$$
\rho_S\!\left(\beta_1,\log_{10}c_1\right)=-0.54,\qquad
\rho_S\!\left(\beta_2,\log_{10}c_2\right)=-0.63,\qquad
\rho_S\!\left(\beta_3,\log_{10}c_3\right)=-0.60.
$$
This posterior geometry is consistent with a trade-off in which higher fitted transmission intensity can be compensated within the model by behavioral adaptation becoming substantial at a lower prevalence. Transmission intensity and the prevalence midpoint of the behavioral response are therefore not sharply separable from aggregate incidence alone.

The wave-specific transmission coefficients themselves were also only moderately differentiated in the Adaptive ($\beta$ variable) specification. Posterior medians and central 95\% intervals were
$$
\beta_1=0.983\;(0.429,\,2.808),\qquad
\beta_2=0.995\;(0.434,\,2.730),\qquad
\beta_3=1.398\;(0.493,\,2.898).
$$
Although the median for $\beta_3$ was higher, the intervals overlapped extensively, and the posterior probabilities
$$
P(\beta_3>\beta_1)=0.640,
\qquad
P(\beta_3>\beta_2)=0.666
$$
provide only limited support for a distinct third-wave transmission regime.

The behavioral delay $\tau$ was similarly weakly resolved. For the Adaptive ($\beta$ global) model, $P(\tau=0)=0.284$, with the remaining posterior mass distributed across delays of one to four weeks. Under the Adaptive ($\beta$ variable) specification, $P(\tau=0)$ increased to 0.417, but most posterior mass still remained distributed across positive delays. The weekly incidence series therefore did not identify a specific behavioral-response delay within the temporal resolution considered here.

Taken together, these results show that the adaptive mechanism can contribute to accurate incidence reconstruction without yielding uniquely determined behavioral parameters. The data provide little evidence for a smooth monotonic evolution of $k_w$ or $c_w$, while the clearest and most reproducible feature of the posterior geometry is the dependence between transmission intensity and the behavioral midpoint. We therefore interpret $k_w$, $c_w$, and $\tau$ as partially identified effective parameters of the incidence-calibrated model rather than as direct measurements of population behavior.

\subsection{Sensitivity analyses}

The main conclusions were broadly robust to the ABC acceptance proportion, although the absolute interval-score values varied with the tolerance. For $q=0.0025$, $0.005$, and $0.01$, respectively, WIS was 947.3, 781.0, and 795.8 for the SIR single-trajectory benchmark; 555.5, 643.8, and 784.9 for the SIR model with wave-specific initialization; 560.5, 653.1, and 779.3 for the Adaptive ($\beta$ global) model; and 711.3, 831.1, and 1030.4 for the Adaptive ($\beta$ variable) model. Across all three acceptance proportions, the SIR model with wave-specific initialization and the Adaptive ($\beta$ global) model remained the two specifications with the lowest WIS. Their differences were small, and their ordering reversed at $q=0.01$, where the adaptive model had a slightly lower WIS. Thus, the primary conclusion that adaptive behavior with shared transmission improves point reconstruction without providing a clear WIS advantage over the simpler wave-specific SIR model was not dependent on the specific choice $q=0.005$.

The Adaptive ($\beta$ variable) specification remained less favorable in terms of interval performance across the tested tolerances, and its WIS increased as the accepted set was broadened. More generally, variation in $q$ changed the absolute performance metrics but did not provide evidence that the additional wave-specific transmission flexibility yielded a consistent advantage over the global-$\beta$ adaptive specification. The sensitivity analysis therefore supports using $q=0.005$ as the primary approximation while interpreting small differences between the wave-specific SIR and Adaptive ($\beta$ global) models cautiously.

A stronger sensitivity was observed when monotonic structure was imposed on the behavioral priors (Table~\ref{tab:ordered_sensitivity}). For the Adaptive ($\beta$ global) model, imposing
$$
k_1\geq k_2\geq k_3,
\qquad
c_1\leq c_2\leq c_3
$$
reduced the mean-based RMSE from 1,038.3 to 876.7, the median-based RMSE from 1,209.6 to 1,051.5, and WIS from 653.1 to 556.3. These correspond to reductions of 15.6\%, 13.1\%, and 14.8\%, respectively. The improved reconstruction therefore shows that constraining the behavioral trajectories can materially regularize the adaptive model.

\begin{table}[htp]
	\centering
	\begin{tabular}{llrrr}
		\toprule
		Model & Behavioral prior & RMSE (mean) & RMSE (median) & WIS \\
		\midrule
		Adaptive ($\beta$ global)
		& Unrestricted
		& 1038.3
		& 1209.6
		& 653.1 \\
		
		Adaptive ($\beta$ global)
		& Ordered
		& 876.7
		& 1051.5
		& 556.3 \\
		
		Adaptive ($\beta$ variable)
		& Unrestricted
		& 2102.6
		& 1067.7
		& 831.1 \\
		
		Adaptive ($\beta$ variable)
		& Ordered
		& 2350.1
		& 774.6
		& 724.9 \\
		\bottomrule
	\end{tabular}
	\caption{Sensitivity of in-sample reconstruction to the behavioral prior structure at the primary ABC acceptance proportion $q=0.005$. Unrestricted specifications use independent wave-specific base priors for $k_w$ and $c_w$. Ordered specifications impose $k_1\geq k_2\geq k_3$ and $c_1\leq c_2\leq c_3$. Lower RMSE and WIS indicate better reconstruction. Because the ordering constraints alter the prior distribution before observing the data, differences between unrestricted and ordered specifications quantify sensitivity to behavioral regularization rather than evidence in favor of monotonic behavioral change.}
	\label{tab:ordered_sensitivity}
\end{table}

The effect of ordering was less uniform for the Adaptive ($\beta$ variable) model. Under unrestricted behavioral priors, its mean-based RMSE, median-based RMSE, and WIS were 2,102.6, 1,067.7, and 831.1, respectively. With ordered priors, the mean-based RMSE increased to 2,350.1, whereas the median-based RMSE decreased to 774.6 and WIS decreased to 724.9. Relative to the unrestricted specification, this represents an 11.8\% increase in mean-based RMSE but reductions of 27.5\% and 12.8\% in median-based RMSE and WIS. The opposing changes in mean- and median-based RMSE again indicate sensitivity to asymmetric posterior-simulation trajectories in the more flexible model.

These improvements under ordered priors should not be interpreted as evidence that the behavioral response actually evolved monotonically across waves. The ordering is imposed before observing the data and changes the wave-specific marginal prior distributions. Moreover, the unrestricted analyses in Section~4.2 provided only limited posterior support for the corresponding complete orderings. The ordered analyses therefore demonstrate that behavioral regularization can materially affect both reconstruction and posterior inference, rather than establishing a data-driven monotonic decline in behavioral sensitivity or increase in response midpoint.

Taken together, the two sensitivity analyses lead to different conclusions. Model-performance comparisons were relatively stable to moderate changes in the ABC acceptance proportion, particularly with respect to the close WIS performance of the wave-specific SIR and Adaptive ($\beta$ global) specifications. In contrast, inference and reconstruction in the adaptive models were meaningfully affected by imposing monotonic behavioral structure. We therefore treat the unrestricted adaptive specifications as the primary analyses and the ordered models as sensitivity analyses illustrating the dependence of behavioral inference on prior structural assumptions.

\section{Discussion and Conclusions}

We compared four SIR-based specifications for reconstructing weekly confirmed
COVID-19 incidence in New York City during June--December 2020, with
particular attention to whether a delayed prevalence-dependent behavioral
response improves reconstruction and whether its parameters can be identified
from aggregate incidence. The results do not support a simple ranking in which
increasing model flexibility consistently improves performance. Instead, they
highlight two related features of multi-wave calibration: substantial gains can
be obtained by allowing different epidemic windows to begin from different
latent states, while the additional decomposition of incidence dynamics into
transmission and behavioral components remains only partially identifiable.

The largest structural improvement occurred when the single continuous SIR
trajectory was replaced by independently initialized epidemic-wave windows.
Relative to the SIR single-trajectory benchmark, the wave-initialized SIR model
reduced the mean- and median-based RMSE by 48.6\% and 53.6\%, respectively,
and reduced WIS by 17.6\% (Table~\ref{tab:measures};
Figure~\ref{fig:ajustes}). This comparison should not, however, be interpreted
as isolating the effect of the initialization parameter alone. The
wave-initialized specification simultaneously partitions the time series and
resets the epidemic state at the beginning of each wave. Its improved
reconstruction therefore indicates that a single uninterrupted SIR trajectory
with fixed transmission and recovery coefficients is too restrictive for the
observed multi-wave pattern, but it does not establish a specific biological
mechanism responsible for the improvement. In this sense, wave-specific
initialization is best viewed as a pragmatic representation of between-wave
dynamics that are absent from the simplified SIR structure.

Adding adaptive behavior while retaining a shared transmission coefficient
produced a more modest and metric-dependent improvement. Relative to the
wave-initialized SIR model, the Adaptive ($\beta$ global) specification reduced
the mean-based RMSE by 23.4\% and the median-based RMSE by 8.8\%, whereas
its WIS was 1.4\% higher. Thus, the adaptive mechanism improved the
reconstruction of the central trajectory but did not provide a clear advantage
in interval performance. This distinction is important because the models were
fitted and evaluated on the same observations and differ substantially in
parameter dimension. The reported RMSE and WIS values therefore quantify
in-sample reconstruction rather than out-of-sample forecasting performance or
complexity-adjusted model preference. Moreover, because the
posterior-simulation intervals arise from ABC parameter uncertainty without a
stochastic observation model, WIS should be interpreted here as a descriptive
measure of parameter-induced interval performance rather than as evidence of
formal probabilistic forecast calibration.

Additional wave-specific flexibility in transmission did not resolve this
trade-off. The Adaptive ($\beta$ variable) model attained the lowest
median-based RMSE among the four specifications, but its mean-based RMSE and
WIS were substantially higher than those of the Adaptive ($\beta$ global)
model (Table~\ref{tab:measures}). The pronounced difference between its
mean- and median-based reconstructions indicates an asymmetric distribution of
posterior-simulated trajectories. Consequently, allowing $\beta$ to vary by wave
does not provide a consistent reconstruction advantage across the criteria
considered here. More generally, these results illustrate that additional
parameter flexibility can improve one summary of fit while simultaneously
increasing posterior uncertainty or asymmetry, particularly when several
parameters can compensate for one another.

This compensation is central to the interpretation of the behavioral
parameters. Across both adaptive specifications, the posterior distributions of
$k_w$ and $c_w$ showed only modest contraction relative to their broad priors
(Figure~\ref{fig:figure2priorposteriorbehavior}). The unrestricted primary
analyses provided little support for a complete monotonic progression in either
behavioral sensitivity or behavioral midpoint. Under the global-$\beta$
specification, the posterior probability of $k_1\geq k_2\geq k_3$ was close to
that expected for a particular ordering under three exchangeable continuous
quantities. Although $c_3$ tended to exceed $c_1$ and $c_2$, its posterior
remained broad. Under wave-specific transmission, the posterior median of
$k_3$ was lower than those of the first two waves, but the corresponding
intervals strongly overlapped and the probability of the complete ordering
remained limited. The behavioral delay $\tau$ was also weakly resolved across
its discrete support. The incidence data therefore do not justify interpreting
the fitted behavioral parameters as precise measurements of changing
population behavior or as evidence for a monotonic decline in behavioral
responsiveness.

The clearest indication of the underlying identifiability problem is instead
the posterior dependence between transmission intensity and the behavioral
midpoint (Figure~\ref{fig:figure3betacconfounding}). Across waves and adaptive
specifications, larger transmission coefficients were systematically associated
with smaller values of $c_w$. Within the model, a larger $\beta$ can therefore
be compensated by a behavioral response that becomes substantial at lower
prevalence. The stronger negative associations observed when transmission was
wave-specific show that introducing additional transmission flexibility does
not automatically separate the two mechanisms. This posterior geometry is
consistent with a broader difficulty in behavioral epidemic models: changes in
aggregate incidence can often be reproduced through different combinations of
intrinsic transmission and feedback from adaptive contact behavior
\cite{eksin2019,weitz2020}. Our analysis illustrates the converse problem as
well: including an explicit behavioral mechanism does not by itself make that
mechanism identifiable from incidence alone.

The fitted coefficients should consequently be interpreted as effective
parameters of the incidence-calibrated system rather than as direct estimates of
biological transmissibility or population risk response. This qualification is
particularly important because confirmed cases are only an ascertained subset
of infections and the model contains no explicit observation process. Changes
over time in testing availability, healthcare-seeking behavior, or the
infection-to-confirmation relationship can therefore be absorbed by $\beta$,
$I_w^{\mathrm{init}}$, or the behavioral parameters. The somewhat higher
posterior median of $\beta_3$ in the wave-specific transmission model is
compatible with greater effective transmission during the final epidemic window,
but the broad and overlapping posterior intervals and only moderate posterior
probabilities for $\beta_3>\beta_1$ and $\beta_3>\beta_2$ do not establish a
distinct third-wave transmission regime. Its interpretation should therefore
remain descriptive rather than biological or causal.

The sensitivity analyses reinforce this cautious interpretation. Varying the
ABC acceptance proportion over the examined range changed the absolute values
of the reconstruction metrics but left the principal comparison relatively
stable: the wave-initialized SIR and Adaptive ($\beta$ global) specifications
remained close in WIS, while the Adaptive ($\beta$ variable) specification did
not acquire a consistent advantage. In contrast, imposing the ordered
behavioral priors $k_1\geq k_2\geq k_3$ and $c_1\leq c_2\leq c_3$ materially
changed reconstruction (Table~\ref{tab:ordered_sensitivity}). For the
global-$\beta$ adaptive model, for example, ordering reduced both RMSE and WIS.
This improvement is evidence that structured behavioral priors can act as
effective regularization, not that the data independently demonstrate monotonic
behavioral change. Because sorting alters the wave-specific prior distributions
before observing the data, posterior monotonicity under this specification is
partly imposed by construction. The contrast between unrestricted and ordered
analyses therefore provides direct evidence that behavioral inference is
sensitive to structural prior assumptions.

Several additional limitations determine the scope of these conclusions.
First, the epidemic-wave boundaries were selected through exploratory inspection
rather than estimated statistically, and all wave-specific results are conditional
on that partition. Independent reinitialization also sacrifices epidemiological
continuity across wave boundaries by resetting the compartmental states and
should be understood as a modeling approximation rather than as a literal
description of population turnover. Second, the deterministic SIR backbone
omits latent infection, asymptomatic transmission, reinfection, heterogeneous
mixing, and other processes that may contribute to the observed trajectory.
Third, weekly aggregation and the integer-valued delay restrict the temporal
resolution at which behavioral feedback can be identified. Fourth, rejection
ABC provides an approximation that depends on the chosen priors, discrepancy,
and acceptance threshold, even though the tolerance sensitivity considered here
suggests that the main model-performance comparison is not driven by the
primary value of $q$. Finally, all performance comparisons are in sample and
contain no explicit adjustment for the substantial differences in model
complexity.

These limitations also indicate where additional information would be most
valuable. In particular, identifiability is unlikely to be solved simply by adding
further flexibility to the incidence model. Auxiliary information on mobility,
testing intensity, seroprevalence, or survey-based precautionary behavior could
provide independent constraints on either the behavioral or observation
processes and thereby reduce the transmission--behavior trade-off. An explicit
observation model linking latent infections to confirmed cases would be
especially useful for separating changes in transmission from changes in
ascertainment. Held-out or cross-validated evaluation would in turn be required
to determine whether the improved in-sample reconstruction of more flexible
specifications translates into predictive benefit. More efficient sequential ABC
methods could facilitate such extensions \cite{toni2009}, although computational
efficiency alone would not resolve structural non-identifiability.

In conclusion, the adaptive-behavior formulation can improve the central
reconstruction of multi-wave COVID-19 incidence relative to a non-adaptive SIR
model with the same wave-specific initialization, but the aggregate incidence
series does not uniquely determine the behavioral response that generated that
improvement. Allowing transmission to vary by wave adds flexibility but does not
produce a consistent gain in reconstruction or resolve the uncertainty in the
behavioral parameters. The most robust inferential result is therefore not a
specific trajectory of behavioral sensitivity, but the evidence of substantial
confounding between transmission intensity and prevalence-dependent contact
adaptation. Multi-wave incidence is compatible with a coupled
behavioral--epidemiological process, but incidence alone is insufficient to
separate its components sharply. Recognizing that limitation is essential when
adaptive epidemic models are used not only to reconstruct observed trajectories,
but also to interpret changes in population behavior.

\bibliographystyle{plain}
\bibliography{references}

@article{kermack1927,
  author  = {Kermack, William Ogilvy and McKendrick, Anderson G.},
  title   = {A Contribution to the Mathematical Theory of Epidemics},
  journal = {Proceedings of the Royal Society of London. Series A},
  volume  = {115},
  number  = {772},
  pages   = {700--721},
  year    = {1927},
  doi     = {10.1098/rspa.1927.0118}
}

@article{hethcote2000,
  author  = {Hethcote, Herbert W.},
  title   = {The Mathematics of Infectious Diseases},
  journal = {SIAM Review},
  volume  = {42},
  number  = {4},
  pages   = {599--653},
  year    = {2000},
  doi     = {10.1137/S0036144500371907}
}

@article{funk2009,
  author  = {Funk, Sebastian and Gilad, Erez and Watkins, Chris and Jansen, Vincent A. A.},
  title   = {The Spread of Awareness and Its Impact on Epidemic Outbreaks},
  journal = {Proceedings of the National Academy of Sciences},
  volume  = {106},
  number  = {16},
  pages   = {6872--6877},
  year    = {2009},
  doi     = {10.1073/pnas.0810762106}
}

@article{funk2010,
  author  = {Funk, Sebastian and Salath{\'e}, Marcel and Jansen, Vincent A. A.},
  title   = {Modelling the Influence of Human Behaviour on the Spread of Infectious
             Diseases: A Review},
  journal = {Journal of the Royal Society Interface},
  volume  = {7},
  number  = {50},
  pages   = {1247--1256},
  year    = {2010},
  doi     = {10.1098/rsif.2010.0142}
}

@article{fenichel2011,
  author  = {Fenichel, Eli P. and Castillo-Chavez, Carlos and Ceddia, M. Graziano
             and Chowell, Gerardo and Gonzalez Parra, Paula A. and Hickling, Graham J.
             and Holloway, Garth and Horan, Richard and Morin, Benjamin
             and Perrings, Charles and Springborn, Michael and Velazquez, Leticia
             and Villalobos, Cristina},
  title   = {Adaptive Human Behavior in Epidemiological Models},
  journal = {Proceedings of the National Academy of Sciences},
  volume  = {108},
  number  = {15},
  pages   = {6306--6311},
  year    = {2011},
  doi     = {10.1073/pnas.1011250108}
}

@article{reluga2010,
  author  = {Reluga, Timothy C.},
  title   = {Game Theory of Social Distancing in Response to an Epidemic},
  journal = {PLOS Computational Biology},
  volume  = {6},
  number  = {5},
  pages   = {e1000793},
  year    = {2010},
  doi     = {10.1371/journal.pcbi.1000793}
}

@article{verelst2016,
  author  = {Verelst, Frederik and Willem, Lander and Beutels, Philippe},
  title   = {Behavioural Change Models for Infectious Disease Transmission:
             A Systematic Review},
  journal = {Journal of the Royal Society Interface},
  volume  = {13},
  number  = {125},
  pages   = {20160820},
  year    = {2016},
  doi     = {10.1098/rsif.2016.0820}
}

@article{perra2021,
  author  = {Perra, Nicola},
  title   = {Non-Pharmaceutical Interventions during the {COVID-19} Pandemic: A Review},
  journal = {Physics Reports},
  volume  = {913},
  pages   = {1--52},
  year    = {2021},
  doi     = {10.1016/j.physrep.2021.02.001}
}

@article{mahmud2025,
  author  = {Mahmud, Md Shahriar and Eshun, Solomon and Espinoza, Baltazar
             and Kadelka, Claus},
  title   = {Adaptive Human Behavior and Delays in Information Availability
             Autonomously Modulate Epidemic Waves},
  journal = {PNAS Nexus},
  volume  = {4},
  number  = {5},
  pages   = {pgaf145},
  year    = {2025},
  doi     = {10.1093/pnasnexus/pgaf145}
}

@article{beaumont2002,
  author  = {Beaumont, Mark A. and Zhang, Wenyang and Balding, David J.},
  title   = {Approximate {Bayesian} Computation in Population Genetics},
  journal = {Genetics},
  volume  = {162},
  number  = {4},
  pages   = {2025--2035},
  year    = {2002},
  doi     = {10.1093/genetics/162.4.2025}
}

@article{toni2009,
  author  = {Toni, Tina and Welch, David and Strelkowa, Natalja and Ipsen, Andreas
             and Stumpf, Michael P. H.},
  title   = {Approximate {Bayesian} Computation Scheme for Parameter Inference
             and Model Selection in Dynamical Systems},
  journal = {Journal of the Royal Society Interface},
  volume  = {6},
  number  = {31},
  pages   = {187--202},
  year    = {2009},
  doi     = {10.1098/rsif.2008.0172}
}

@article{csillery2010,
  author  = {Csill{\'e}ry, Katalin and Blum, Michael G. B. and Gaggiotti, Oscar E.
             and Fran{\c{c}}ois, Olivier},
  title   = {Approximate {Bayesian} Computation ({ABC}) in Practice},
  journal = {Trends in Ecology \& Evolution},
  volume  = {25},
  number  = {7},
  pages   = {410--418},
  year    = {2010},
  doi     = {10.1016/j.tree.2010.04.001}
}

@article{gneiting2007,
  author  = {Gneiting, Tilmann and Raftery, Adrian E.},
  title   = {Strictly Proper Scoring Rules, Prediction, and Estimation},
  journal = {Journal of the American Statistical Association},
  volume  = {102},
  number  = {477},
  pages   = {359--378},
  year    = {2007},
  doi     = {10.1198/016214506000001437}
}

@article{bracher2021,
  author  = {Bracher, Johannes and Ray, Evan L. and Gneiting, Tilmann
             and Reich, Nicholas G.},
  title   = {Evaluating Epidemic Forecasts in an Interval Format},
  journal = {PLOS Computational Biology},
  volume  = {17},
  number  = {2},
  pages   = {e1008618},
  year    = {2021},
  doi     = {10.1371/journal.pcbi.1008618}
}

@misc{nycdata2025a,
  author       = {{New York City Department of Health and Mental Hygiene}},
  title        = {{COVID-19}: Data Trends and Totals},
  year         = {2025},
  howpublished = {\url{https://www.nyc.gov/site/doh/covid/covid-19-data-totals.page}},
  note         = {Accessed 2026-04-17}
}

@misc{nycdata2025b,
  author       = {{New York City Department of Health and Mental Hygiene}},
  title        = {\texttt{data-by-day.csv}},
  year         = {2025},
  howpublished = {GitHub repository \texttt{nychealth/coronavirus-data},
                  file \texttt{trends/data-by-day.csv}.
                  \url{https://github.com/nychealth/coronavirus-data}},
  note         = {Accessed 2026-04-17}
}

@misc{reuters2020,
  author       = {{Reuters}},
  title        = {{New York City} to Enter Phase Two of Reopening on {June} 22 --- Mayor},
  year         = {2020},
  month        = jun,
  howpublished = {\url{https://www.reuters.com/article/health-coronavirus-usa-new-york-city/new-york-city-to-enter-phase-two-of-reopening-on-june-22-mayor-idUKL1N2DU165/}},
  note         = {Reporting by Maria Caspani; accessed 2026-04-17}
}

@misc{lambert2020,
  author       = {Lambert, Lisa and Borter, Gabriella and Allen, Jonathan},
  title        = {{`Race Against Time'}: First {Americans} Vaccinated as {U.S.}
                  Death Toll Passes 300,000},
  year         = {2020},
  month        = dec,
  howpublished = {Reuters.
                  \url{https://www.reuters.com/business/healthcare-pharmaceuticals/race-against-time-first-americans-vaccinated-us-death-toll-passes-300000-2020-12-14/}},
  note         = {Updated December 15, 2020; accessed 2026-04-17}
}

@article{eksin2019,
  author  = {Eksin, Ceyhun and Paarporn, Keith and Weitz, Joshua S.},
  title   = {Systematic Biases in Disease Forecasting: The Role of Behavior Change},
  journal = {Epidemics},
  volume  = {27},
  pages   = {96--105},
  year    = {2019},
  doi     = {10.1016/j.epidem.2019.02.004}
}

@article{weitz2020,
  author  = {Weitz, Joshua S. and Park, Sang Woo and Eksin, Ceyhun
             and Dushoff, Jonathan},
  title   = {Awareness-Driven Behavior Changes Can Shift the Shape of Epidemics
             Away from Peaks and Toward Plateaus, Shoulders, and Oscillations},
  journal = {Proceedings of the National Academy of Sciences},
  volume  = {117},
  number  = {51},
  pages   = {32764--32771},
  year    = {2020},
  doi     = {10.1073/pnas.2009911117}
}

\appendix
\clearpage
\section{Supplementary Material}

\begin{table}[htp]
	\centering
	\begin{tabular}{lcc}
		\toprule
		Posterior event
		& Adaptive ($\beta$ global)
		& Adaptive ($\beta$ variable) \\
		\midrule
		$P(k_1 \geq k_2)$
		& 0.499 & 0.508 \\
		
		$P(k_1 \geq k_3)$
		& 0.481 & 0.618 \\
		
		$P(k_2 \geq k_3)$
		& 0.487 & 0.627 \\
		
		$P(k_1 \geq k_2 \geq k_3)$
		& 0.183 & 0.250 \\
		
		\addlinespace
		
		$P(c_1 \leq c_2)$
		& 0.510 & 0.513 \\
		
		$P(c_1 \leq c_3)$
		& 0.770 & 0.629 \\
		
		$P(c_2 \leq c_3)$
		& 0.772 & 0.631 \\
		
		$P(c_1 \leq c_2 \leq c_3)$
		& 0.329 & 0.249 \\
		\bottomrule
	\end{tabular}
	
	\caption{Posterior probabilities of pairwise and joint ordering of the
		wave-specific behavioral parameters under the two unrestricted adaptive
		model specifications. Probabilities were computed from the 2,500
		ABC-accepted draws at $q=0.005$. The wave-specific behavioral parameters
		were assigned independent exchangeable base priors in the primary analyses,
		so no monotonic ordering was imposed a priori.}
	
	\label{tab:supp_ordering_probabilities}
\end{table}

\end{document}